
\input harvmac
\overfullrule=0pt
\abovedisplayskip=12pt plus 3pt minus 3pt
\belowdisplayskip=12pt plus 3pt minus 3pt
\sequentialequations
%

\def\bar{\overline}
\def\to{\rightarrow}

\def\btau{{\bar\tau}}
\def\barL{{\bar L}}
\def\barc{{\bar c}}
\def\barq{{\bar q}}

\font\zfont = cmss10 
\font\litfont = cmr6

\def\bigone{\hbox{1\kern -.23em {\rm l}}}
\def\ZZ{\hbox{\zfont Z\kern-.4emZ}}
\def\half{{\litfont {1 \over 2}}}
\def\sltz{$SL(2,Z)$}


\lref\beckers{K. Becker and M. Becker, {\it ``M-Theory on Eight
Manifolds''}, Nucl. Phys. {\bf B477} (1996) 155; hep-th/9605053.} 
\lref\svw{S. Sethi, C. Vafa and E. Witten, {\it ``Constraints on
Low-Dimensional String Compactifications''}, 
Nucl. Phys. {\bf B480} (1996) 213; hep-th/9606122.}
\lref\dmlow{K. Dasgupta and S. Mukhi, {\it ``A Note on Low-Dimensional
String Compactification''}, Phys. Lett. {\bf B398} (1997) 285; 
hep-th/9612188.}
\lref\vafwit{C. Vafa and E. Witten, {\it ``A One Loop Test Of String
Duality''}, Nucl. Phys. {\bf B447} (1995) 261; hep-th/9505053.}
\lref\dlm{M.J. Duff, J.T. Liu and R. Minasian, {\it 
``Eleven-dimensional Origin of String/String Duality: a 
One Loop Test''}, Nucl. Phys. {\bf B452} (1995) 261; hep-th/9506126.}
\lref\shibaji{S. Roy, {\it ``An Orbifold and Orientifold of Type IIB
Theory on K3 $\times$ K3''}; hep-th/9607157.}
\lref\ganor{O. Ganor, {\it ``Compactification of Tensionless 
String Theories''}; hep-th/9607092.}
\lref\vafaf{C. Vafa, {\it ``Evidence for F-theory''}, Nucl. Phys. {\bf
B469} (1996) 403; hep-th/9602022.}
\lref\greengab{M. Gaberdiel and M.B. Green, {\it ``An \sltz\ anomaly in IIB
Supergravity and its F-theory Interpretation''}; hep-th/9810153.}
\lref\alwit{L. Alvarez-Gaum\'e and E. Witten, {\it ``Gravitational
Anomalies''}, Nucl. Phys. {\bf B234} (1984), 269.}
\lref\witflux{E. Witten, {\it ``On Flux Quantization in M-theory and the 
Effective Action''}, J. Geom. Phys. {\bf 22} (1997) 1; hep-th/9609122.} 
\lref\djm{K. Dasgupta, D.P. Jatkar and S. Mukhi,
{\it ``Gravitational Couplings and $Z_2$ Orientifolds''}, 
Nucl. Phys. {\bf B523} (1998) 465; hep-th/9707224.}
\lref\hennteit{M. Henneaux and C. Teitelboim, {\it ``Dynamics of Chiral
(Selfdual) $p$-forms''}, Phys. Lett. {\bf B206} (1988)
650.} 
\lref\dasmfive{K. Dasgupta and S. Mukhi, {\it ``Orbifolds of M-theory''},
Nucl. Phys. {\bf B465} (1996) 399; hep-th/9512196.}
\lref\witmfive{E. Witten, {\it ``Five branes and M-theory on an
Orbifold''}, Nucl. Phys. {\bf B463} (1996) 383; hep-th/9512219.}
\lref\bsv{M. Bershadsky, V. Sadov and C. Vafa, {\it ``D-branes and
Topological Field Theories''}, 
Nucl. Phys. {\bf B463} (1996) 420; hep-th/9611222.}
\lref\ghm{M.B. Green, J.A. Harvey and G. Moore, {\it ``I-brane Inflow and 
Anomalous Couplings on D-branes''}, 
Class. Quant. Grav. {\bf 14} (1997) 47; hep-th/9605033.}
\lref\dminf{K. Dasgupta and S. Mukhi, {\it ``Anomaly Inflow on Orientifold
Planes''}, JHEP {\bf 03} (1998) 004; hep-th/9709219.}
\lref\crapsroose{B. Craps and F. Roose, {\it ``Anomalous D-brane and
Orientifold Couplings from the Boundary State''}; hep-th/9808074.} 
\lref\afmn{I. Antoniadis, S. Ferrara, R. Minasian and 
K.S. Narain, {\it ``$R^4$ Couplings in M and Type II Theories on 
Calabi-Yau Spaces''}, Nucl. Phys. {\bf B507} (1997) 571; 
hep-th/9707013.} 

{\nopagenumbers
\Title{\vtop{\hbox{hep-th/9810213}
\hbox{HUTP-98/A075}
\hbox{TIFR/TH/98-41}}}
{\centerline{Dualities and the \sltz\ Anomaly}}
\centerline{Sunil Mukhi\foot{E-mail: mukhi@tifr.res.in}}
\vskip 6pt
\centerline{\it Lyman Laboratory of Physics, Harvard University}
\centerline{\it Cambridge, MA 02138, U.S.A.}
\vskip 3pt
\centerline{and}
\vskip 3pt
\centerline{\it Tata Institute of Fundamental Research,}
\centerline{\it Homi Bhabha Rd, Bombay 400 005, India}
\ \medskip
\centerline{ABSTRACT}

The \sltz\ anomaly recently derived for type IIB supergravity in 10
dimensions is shown to be a consequence of T-duality with the type IIA
string, after compactification to 2 dimensions on an 8-fold. This explains
the identity of the gravitational 8-forms appearing in different contexts
in the effective actions of type IIA and IIB string theories. In this
framework, constraints on the compactification manifold arise from modular
invariance of the 2d theory. Related issues in 6 dimensions are examined,
and it is argued that similar anomalies are present on the worldvolumes of
M-theory 5-branes and orientifold 5-planes.

\ \vfill 
\leftline{October 1998}
\eject} 
\ftno=0
\newsec{Introduction} 

It has recently been observed\refs\greengab\ that a certain chiral $U(1)$
transformation on fermions in 10d type IIB supergravity is anomalous. This
anomaly is not irreducible, but rather can be cancelled by the addition of
a local counterterm. The counterterm fails to be invariant under \sltz\ 
duality transformations of the theory, hence this result is interpreted as
an \sltz\  anomaly. It turns out that the counterterm leading to this
anomaly vanishes when evaluated on most ``standard'' compactifications of
type IIB, since it depends on derivatives of the dilaton and axion fields
over the compactification manifold. Hence it turns out to give constraints
only on compactifications of ``F-theory''\refs\vafaf\ type.

The anomaly is proportional to a certain 8-form built out of Riemann
curvature 2-forms. This turns out to be precisely the form whose integral
over an 8-manifold gives the Euler characteristic of the manifold. The same
8-form appears in the low-energy Lagrangians of type IIA string
theory\refs\vafwit\ as well as M-theory\refs\dlm, but has apparently not
appeared so far in the Lagrangian of type IIB string theory.  In
particular, its appearance in the present context seems to be a remarkable
coincidence.

One purpose of this note is to provide an alternate derivation of the
\sltz\ anomaly using T-duality and freedom from gravitational anomalies. 
Among other things, this provides an explanation for the apparent
coincidence noted above. 

T-duality between type IIA and IIB string theories compactified on at least
one spatial circle, requires the presence of certain couplings on
compactified type IIB, which, however, are absent in the uncompactified
theory. These couplings lead to tadpoles in naive low-dimensional
compactifications of the type IIA theory\refs\svw. It was explained in
Ref.\dmlow\ that this strange phenomenon in type IIB arises from the
chirality of the theory, which is preserved by its 8-fold compactifications
down to 2 spacetime dimensions. This chirality induces a ``vacuum
momentum'' (first noted in a similar context in Ref.\ganor) which is
related by T-duality to the type IIA tadpole, or ``vacuum winding
charge''. In Ref.\dmlow, the spectrum of compactified IIB theory was
explicitly computed and shown to reproduce the expected vacuum momentum.

All this is related to the \sltz\ anomaly as follows. Suppose we compactify
the type IIB string on an 8-fold down to 2 dimensions. Because the
resulting theory is chiral, the anomalous $U(1)$ of Ref.\greengab\
potentially survives compactification. As will be shown below, a linear
relation holds between this $U(1)$ anomaly, the two-dimensional
gravitational anomaly, and the vacuum momentum in 2d. Although potentially
the 2d chiral theory has gravitational anomalies, these were shown to
cancel in Ref.\dmlow. It follows that the $U(1)$ anomaly is equated to the
vacuum momentum of the theory.

This vacuum momentum is determined by T-duality with type IIA (and
independently by the computation in Ref.\dmlow) to be proportional to
$\chi/24$ where $\chi$ is the Euler characteristic of the 8-fold. The
result is that the \sltz\ anomaly is completely predicted by this argument,
which provides the link to type IIA and explains the ubiquitous role of
the Euler 8-form.

In this picture, the constraint $\chi/24\in Z$ is a consequence of
modular invariance of the 2d theory. Moreover, it will be argued that this
constraint is weakened if we allow for background fluxes, much as in type
IIA- and M-theory. This in turn gives rise to a prediction that the anomaly
computed in Ref.\greengab\ will be modified if the 4-form potential $D^+$
is nonzero. 

The relationship between anomalies and vacuum momentum that is studied here
in type IIB theory also makes a fascinating appearance in M-theory. This
happens via the fact, which will be argued below, that M-theory 5-branes
and orientifold 5-planes have analogous anomalies on their world volumes
and give rise to vacuum momentum when wrapped on suitable 4-folds.

Throughout this paper, the principal focus will be on the anomaly in the
chiral $U(1)$ symmetry described in Ref.\greengab. From a string theory
point of view, it may be debated whether this anomaly really needs to be
cancelled by a local counterterm as is done in Ref.\greengab, since the
global $SL(2,R)$ and local $U(1)$ symmetry relevant in that discussion
seem to be more a feature of type IIB supergravity than of string
theory. If we choose to cancel it then there is an $SL(2,Z)$ anomaly, but
if we do not, the $U(1)$ anomaly remains. In either case, it is the
surviving anomaly that is related by duality, as described below, 
to a number of interesting phenomena.

\newsec{Type IIB on 8-folds}

The relevant results in Ref.\dmlow\ for type IIB compactifications on
8-folds will first be reviewed briefly.

On compactifying type IIB on a circle, it becomes equivalent to IIA on a
circle under T-duality. Thus it must possess the dimensional reductions of
both the classical term $\int B\wedge dC \wedge dC$ and the one-loop term
$\int B \wedge I_8$ in type IIA theory (here $B$ is the NS-NS 2-form and C
is the RR 3-form of type IIA, while $I_8(R)$ is a polynomial in the
curvature, defined below). After reducing on a circle, the
$B$-field of type IIA becomes a 1-form $A$ which measures the winding
charge with respect to that circle. Under T-duality this, in turn, becomes
the Kaluza-Klein 1-form arising by reduction of the 10D metric of IIB on
the circle. Thus we must look for terms in type IIB which reduce to $\int
A\wedge dC \wedge dC$ and $A\wedge I_8$ in 9 dimensions.

The origin of the tree-level term is explained in Ref.\dmlow, and will not
be relevant here. The one-loop term is far more subtle. It is known that in
10 dimensions there is no one-loop correction in type IIB analogous to the
term $\int B\wedge I_8$ in type IIA. Moreover, one can easily convince
oneself directly that there is no purely gravitational term that one can
write down in 10d which reduces to $\int A\wedge I_8$ in 9d, with $A$ being
the KK gauge field. In fact, it turns out that no modification is required
in 10d to the type IIB action, but as soon as one compactifies on a circle,
however large, there is a radius-dependent term of the desired form in
9d.

Suppose we compactify both type IIA and IIB on the same 8-fold and then
further on a circle to $0+1$ dimensions. T-dualizing along the circle maps
one theory to the other. Now we have an apparent puzzle: type IIA has a
2-form tadpole in 2d\refs\svw, which will become a 1-form tadpole in 1d,
and this is proportional to the Euler characteristic $\chi$ of the
eightfold. However, there is no inconsistency for type IIB on the eightfold
to two dimensions (for example, gravitational anomalies cancel, as was
demonstrated in Ref.\dmlow), so the inconsistency required by T-duality
must arise upon compactifying one further dimension. Moreover, it must take
the form of a tadpole for the KK 1-form $A=g_{12}$.

For 2d field theories on a cylinder, the generator of translations along
the compact direction is $L_0 - {\bar L}_0$. Thus, a nonzero value of this
operator in the vacuum implies that, from a 2d point of view, there is a
nonzero momentum in the vacuum state. Under T-duality, this will turn into
a nonzero winding charge of the vacuum, just what we would expect in a
theory which has a 2-form tadpole in 2d. The tadpole must have the precise
value $\chi/24$ (for the special case of $K3\times K3$, a similar
argument was given by Ganor\refs\ganor).

Explicit computation in the compactified type IIB theory indeed
shows\refs\dmlow\ that
\eqn\lolobar{
(L_0 - {\bar L}_0)_{\rm vac} = {\chi\over 24} }
as predicted by T-duality. 

Note that if the circle becomes large and we are effectively in two
noncompact dimensions, this effect goes away. The reason is that the
operator $L_0$ as conventionally defined in conformal field theory has a
zero-point contribution $-{1\over 24}$ for a free boson only if the radius
of the circle (the range of the $\sigma$ coordinate) is fixed to be
$2\pi$. For a circle of radius $2\pi R$, the zero-point contribution is
actually $-{1\over 24 R}$, so that it goes away in the limit
$R\to\infty$. This explains why there is no corresponding one-loop term in
the effective action of type IIB theory in 2 (or 6 or 10) dimensions, and
yet the prediction of T-duality with type IIA is satisfied.

\newsec{The \sltz\ Anomaly}

In what follows, we will always consider spin manifolds with at least one
non-vanishing spinor.

Let us define the 8-form 
\eqn\xeight{
I_8(R) = -{1\over (2\pi)^4}{1\over 8}\left(\tr R^4 - {1\over 4}(\tr R^2)^2
\right) }
which has the property that
\eqn\inti{
\int_{M^8} I_8(R) = \chi }
where $\chi$ is the Euler characteristic of the 8-manifold $M_8$.

We can also define another 8-form, the signature
8-form:
\eqn\signeight{
J_8(R) = -{1\over (2\pi)^4}{1\over 180}\left(7\, \tr R^4 - {5\over 2}
(\tr R^2)^2 \right)}
which satisfies
\eqn\intj{
\int_{M^8} J_8(R) = \tau }
where $\tau$ is the signature of the 8-manifold $M_8$.

It is evident that there are precisely two independent 8-forms that one can
make out of traces of products of the Riemann 2-form, these can be
parametrized as $\tr R^4$ and $(\tr R^2)^2$, or as $p_2$ and $(p_1)^2$
($p_i$ are the Pontryagin classes), or as $I_8(R)$ and $J_8(R)$.

Now, on general grounds we can assume the $U(1)$ anomaly in 10d type IIB
supergravity to be given by 
\eqn\anom{
\Delta = - \int {F\over 4\pi}\wedge A_8(R)\, \Sigma(x)}
Here, $\Sigma(x)$ is the
parameter of the $U(1)$ transformation, and $A_8(R)$ is an unknown 8-form
which can, of course, be parametrized as a linear combination of $I_8(R)$
and $J_8(R)$ that we have defined above. The 2-form $F$ is defined in terms
of the complex dilaton-axion field $\tau$ as
\eqn\effdef{
F= i{d\btau\wedge d\tau\over 4(\tau_2)^2}}

The form of Eq.\anom\ follows from
the fact that (i) there is no nonzero 10-form made entirely from traces of
products of $R$, (ii) the 2-form $F$ satisfies $F\wedge F=0$ by virtue of
its definition.

With this expression for the $U(1)$ anomaly, there is a local counterterm
\eqn\localcount{
\int \phi\, {F\over 4\pi} \wedge A_8(R)}
where $\phi$ is a scalar field which is pure gauge, corresponding to the
$U(1)$ part of the $SL(2,R)$ variables. Under the gauge transformation
$\delta\phi = \Sigma(x)$, the variation of Eq.\localcount\ cancels the
anomaly in Eq.\anom.

In Ref.\refs\greengab, it is shown by explicit computation that  
\eqn\explanom{
A_8(R) = {1\over 6} I_8(R) }
In the following, it will be shown that this is a consequence of the vacuum
momentum predicted by T-duality with the type IIA string, and confirmed in
Ref.\dmlow.

\newsec{The $U(1)$ Anomaly and Vacuum Momentum} 

Consider compactifying the type IIB string on a (spin) 8-fold $M^8$ to 2
spacetime dimensions.  The resulting 2-dimensional theory is chiral, in
fact it has $(0,2)$ chiral supersymmetry for 8-folds with $spin(7)$
holonomy, $(0,4)$ for $SU(4)$ holonomy, $(0,8)$ for $SU(2)\times SU(2)$
holonomy (the case of $K3\times K3$) and $(0,16)$ if the holonomy is
contained in $SU(2)$, which is true for the orbifold $T^8/Z_2$. 
The spectrum for all these cases is worked out in Ref.\dmlow, but we will
not need this for the following argument.

A generic 2d supergravity theory with $(0,N)$ supersymmetry coupled to
matter has the following spectrum:
\eqn\genspect{
\eqalign{
1~ {\rm supergravity~ multiplet} &:\qquad (g_{\mu\nu},
\phi, N\psi_\mu^-, N\psi^+)\cr
\left({n_+\over N}\right)~ 
{\rm chiral~ multiplets} &:\qquad (N\phi^+, N\psi^+)\cr
n_-^\phi~ {\rm anti-chiral~ scalars} &:\qquad \phi^-\cr
n_-^\psi~ {\rm anti-chiral~ fermions} &:\qquad \psi^- \cr }}
Note that the anti-chiral matter fields are supersymmetry singlets.

Now, purely in terms of the integers $N,n_+,n_-^\phi,n_-^\psi$, we can
compute various interesting quantities in this theory. For now,
we are not using the fact that this 2d supergravity theory comes from type
IIB or any other compactification. 

Suppose this 2d theory has a chiral $U(1)$ symmetry under which the
gravitinos have charge $\alpha$ and the other fermions have charge
$\beta$. The anomaly in this symmetry can be written:
\eqn\twoduone{
\Delta = - 4 A_{U(1)}^{\alpha,\beta}\int F\;\Sigma(x)}
where $F$ is the $U(1)$ field strength. (In the present case the $U(1)$
gauge field is composite, and $F$ is as given in Eq.\effdef.) 
The coefficient of the anomaly is given by\refs\alwit:
\eqn\uonecoeff{
A_{U(1)}^{\alpha,\beta} = 
\alpha N + {\beta}{1\over 24} (n_+ - n_-^\psi) }
This theory could also have a gravitational anomaly, for which the
coefficient would be proportional to
\eqn\gravcoeff{
A_{\rm grav}= \half N + 
{n_+ - n_-^\phi \over 24} + {n_+ - n_-^\psi \over 48} }
Finally, the chiral theory could in principle have a non-vanishing
zero-point energy. Supersymmetry actually sets this to zero for chiral
fields, which lie in supermultiplets, but does not determine it for the
anti-chiral fields which are supersymmetry singlets. We have:
\eqn\vacmom{
A_{\rm vac}\equiv (L_0 - \barL_0)_{\rm vac} = {n_-^\phi -
n_-^\psi\over 24 }}
We can now specialize to the case where the chiral $U(1)$ has charges
$\half,{3\over 2}$ on the gravitino and spin-$\half$ fermions respectively.
From the above three equations, we easily read off the relation
\eqn\reln{
A_{U(1)}^{\half,{3\over 2}} = A_{\rm grav} + A_{\rm vac} }
It follows that if the 2d theory is free of gravitational anomalies, then
the $U(1)$ anomaly (which leads to an \sltz\ anomaly by the arguments of
Ref.\greengab) is numerically equal to the vacuum momentum. 

As a check, we use formulae from Ref.\dmlow\ for 8-fold compactifications
of the type IIB string, and find
\eqn\insertform{
\eqalign{
A_{U(1)}^{\half,{3\over 2}} &= {\chi\over 24}\cr
A_{\rm grav} &= 0\cr
A_{\rm vac} &= {\chi\over 24}\cr }}
To summarize, we have shown that vacuum momentum, itself a consequence of
T-duality with the type IIA string, is the origin of the $U(1)$ anomaly in
2d. Now, because there are only two independent gravitational 8-forms, the
corresponding $U(1)$ anomaly in 10d is also uniquely determined to be the
one in Eqs.\anom, \explanom, as found in Ref.\greengab.

\newsec{Constraints on Compactifications}

In Ref.\greengab, the presence of a $U(1)$ anomaly leads to a counterterm
which in turn induces an \sltz\ anomaly. The consistency requirements
discussed in Ref.\refs\greengab\ imply that $\chi/24\in Z$. As we will now
see, the same constraint arises very naturally in the present framework by
requiring modular invariance of the 2d theory.

Viewed as a chiral 2-dimensional CFT, the theory obtained by
compactifying the type IIB string on an 8-fold has a partition function:
\eqn\partfn{
Z= \tr\, q^{L_0- {c\over 24}} \barq^{\barL_0- {\barc\over 24}}}
Under a modular transformation $\tau\rightarrow \tau +1$, the partition
function is invariant only if every state $|\Phi\rangle$ in the Hilbert
space of the theory satisfies
\eqn\invif{
e^{2\pi i \left( (L_0 - \barL_0) - (c-\barc)/24 \right)}|\Phi\rangle 
= |\Phi\rangle }
Thus the exponent is required to be an integer. 
In particular, the vacuum state should satisfy this requirement, from
which we get:
\eqn\vacreq{
\left( (L_0 - \barL_0)_{\rm vac} - (c-\barc)/24 \right)\in Z }
The left hand side is precisely the sum of the vacuum momentum and the
gravitational anomaly, $A_{\rm vac} + A_{\rm grav}$. In 8-fold
compactifications of type IIB, this quantity turns out to be ${\chi\over
24}$, as we have seen, so quantization of this number follows very
naturally from modular invariance in 2d.

It should be noted that the constraint $\chi/24\in Z$ only applies
to an elementary class of compactifications {\it without fluxes}. It is
known, for example, that in M-theory and type IIA string theory,
supersymmetric compactifications on 8-folds can include nonzero values of
$\int dC\wedge dC$ over the 8-fold\refs\beckers. Later it was
shown\refs\witflux\ that $dC$ is actually allowed to have half-integral
flux over a 4-cycle, and that consistent compactifications can be defined
on 8-manifolds for which $\chi$ is only a multiple of 6, but not of 24, so
long as there is a suitable nonzero value of $\int dC\wedge dC$ in the
background.

It has been analogously argued that various other field strengths can have
fractional flux in string theory\refs\djm, but a consistent
compactification of type IIB on 8-folds with non-integer $\chi/24$ does not
appear to have been constructed. For F-theory, fluxes are irrelevant to
this particular question, because elliptically fibred Calabi-Yau complex
4-folds anyway have integer $\chi/24$\refs\svw.

Starting from the fact that fractional 4-form field-strengths $dC$ must be
turned on in M-theory and type IIA, for manifolds not satisfying
$\chi/24\in Z$, it will now be argued that the formulae for the
$U(1)$ and $SL(2,Z)$ anomalies in Ref.\refs\greengab\ necessarily receive
corrections when the 5-form field strength $dD^+$ of type IIB is nonzero
(here $D^+$ stands for the self-dual 4-form potential). However, these
corrections cannot apparently be written in Lorentz-covariant form, which
is a manifestation of the well-known impossibility of writing
a covariant action when $D^+$ is nonzero.

The argument goes as follows. Suppose we pick $M^8$ to be a Calabi-Yau
complex 4-fold with $\chi$ a multiple of 6 but not of 24. Now compactify
type IIA theory on this down to 2 dimensions, and include a half-integral
flux $dC$ over 4-cycles so that 
\eqn\chiflux{
{\chi\over 24} - {1\over 8\pi^2} \int dC\wedge dC \in Z }
If this quantity is moreover positive, it can be cancelled
by including the right number of type IIA strings in the vacuum and we have
a tadpole-free compactification (the relevant equation, correcting a sign
in \refs\beckers\ and incorporating the vacuum branes of Ref.\refs\svw, can
be found in Ref.\refs\dmlow). But for the moment we do not include these
branes, and just consider this compactification to 2d, with its associated
tadpole problem. 

Compactifying further on a circle and T-dualizing implies that the T-dual
type IIB has a vacuum momentum
\eqn\vacwflux{
(L_0 - \barL_0)_{\rm vac} = {\chi\over 24} - 
{1\over 8\pi^2} \int dC\wedge dC \in Z }
What is $dC$ from the type IIB point of view? If we label the circle
direction as $x^9$ and let the coordinates of the 8-fold be
$x^i$, $i= 1,\ldots, 8$ then we have the T-duality relation:
\eqn\tdualreln{
C_{ijk} = D^+_{ijk1}}
from which it follows that
\eqn\fluxreln{
\eqalign{
\int_{M^8}dC\wedge dC &\sim \int_{M^8} \epsilon^{ijkli'j'k'l'}
(dD^+)_{ijkl1}\wedge (dD^+)_{i'j'k'l'1}\cr
&\sim (dD^+)_{ijkl0}(dD^+)^{ijkl}_{~~~~1}\cr}}
where the last equality follows from self-duality of $dD^+$ in 10
dimensions.  This is of course not manifestly covariant under $SO(9,1)$,
but only under the subgroup $SO(8)\times SO(1,1)$.

Tracing back the relationship discussed in the previous sections, we find
that the formula for the $U(1)$ anomaly in 10 dimensions must be modified
as follows:
\eqn\modanom{
\Delta = -  \int {F\over 4\pi}\wedge {I_8(R)\over 6}\, \Sigma(x)\rightarrow
-\int {F\over 4\pi}\wedge 4 \left({I_8(R)\over 24}- {1\over 8\pi^2}
(dD^+)_1\wedge (dD^+)_1\right) \Sigma(x)}
where the extra term is shorthand for the last term in
Eq.\fluxreln. Perhaps it is possible to write this is in a better
way, as in Ref.\refs\hennteit, but that direction will not be pursued
here. The main point is to note that the $SL(2,Z)$ anomaly does not
necessarily imply integer quantization of $\chi/24$, but that
background fluxes will provide an ``escape route'' from this rule,
as in type IIA and M-theory.

\newsec{Relation to M-theory 5-branes and Orientifold 5-planes}

In this section we note that the above considerations have some
implications for 5-branes and orientifold 5-planes of M-theory. 

Suppose we compactify type IIB string theory to 6 dimensions on K3. The
10-dimensional $U(1)$ anomaly of Ref.\refs\greengab\ descends to a
6-dimensional $U(1)$ anomaly by simply using $p_1(K3)=48$, from which one
finds:
\eqn\sixdanom{
\Delta = 2\int {F\over 4\pi}\wedge p_1(R)\, \Sigma(x)}
Clearly, a further compactifiction on $K3$ will reproduce the results above,
for the special case where the 8-fold is $K3\times K3$. Instead of
doing that, we first use a nontrivial duality that relates the present
model to the orientifold of M-theory on 
$T^5/Z_2$\refs\dasmfive\refs\witmfive. This compactification has 16
M-theory 5-branes and 32 orientifold 5-planes in the vacuum. These are the
only chiral objects in the theory; the bulk is 11-dimensional M-theory and
hence non-chiral. 

Now we ask what is the interpretation of the above anomaly in terms of
the M-theory defects. Because the bulk is non-chiral, one must assume
that the anomaly comes from branes and/or planes. It turns out that
this involves an interesting phenomenon about M-branes and
planes\foot{Related issues were discussed in Ref.\refs\afmn. 
Some of the observations below emerged in subsequent discussions
with K.S. Narain.}. 

It is well-known that D-branes in type II string theories carry WZ-type
gravitational couplings on their world-volumes\refs\bsv\refs\ghm, and more
recently it has been noted that orientifold planes in the same theories
also carry localized gravitational WZ
couplings\refs\djm\refs\dminf. However, the question of existence of
gravitational couplings on defects in M-theory has not so far been
answered, though it was raised at the end of a recent
paper\refs\crapsroose.

Actually it is quite easy to see that the M-theory 5-brane indeed cannot
have gravitational WZ couplings on its world-volume (such couplings would
have to be proportional to the 4-form $p_1(R)$ and another spacetime
2-form, but the latter does not exist in uncompactified M-theory). However,
once we compactify M-theory on a circle and wrap the 5-brane on this
circle, we obtain the D 4-brane of type IIA theory and this certainly has a
gravitational coupling $\int A\wedge p_1(R)$ on its world-volume, where $A$
is the Ramond-Ramond 1-form of type IIA theory or the Kaluza-Klein gauge
field of compactified M-theory. Thus we have a puzzle: how does the M
5-brane produce this term when wrapped on a circle given that it does not
have it to start with? The same question can also be asked about the
orientifold 5-plane, since after wrapping, it is dual to the orientifold
4-plane of type IIA, which again has gravitational WZ couplings\refs\djm.

The question is quite analogous to the one asked in Ref.\dmlow\ about the
origin of $\int A\wedge I_8(R)$ in type IIB theory compactified on a circle
to 9d, given that no corresponding term is present in 10d. The answer also
turns out to be analogous, and fits beautifully with the $SL(2,Z)$ anomaly
in Eq.\sixdanom\ above.

The key point is that, like the type IIB theory in 10d, the M 5-brane and M
5-plane are chiral objects. Hence, when they wrap on a suitable 4-fold, the
resulting 2d theories are still chiral, indeed if the 4-fold is $K3$ then
one gets theories with chiral supersymmetry in 2d. These theories can
possess a vacuum momentum, which after compactification of one more
dimension, turns into the expected tadpole predicted by the relation to
D-branes and string theory orientifold planes.

This in turn means that the 6-dimensional field theories on the 5-brane and
orientifold 5-plane have $U(1)$ anomalies. In fact, in this case one
expects a multiplet of anomalies, since the duality group is $SO(5,21,Z)$,
much larger than $SL(2,Z)$ of ten-dimensional type IIB. The details need
not be spelled out here since they follow in a straightforward way from
analogous considerations elsewhere in this paper and in Ref.\djm. The
result is that the $U(1)$ under which gravitinos have charge $\half$ and
the remaining fermions have charge ${3\over 2}$ is anomalous on an M-theory
5-brane, the anomaly being
\eqn\fivebanom{
\Delta = {1\over 12}\int_{M^6} {F\over 4\pi}\wedge p_1(R)\, \Sigma(x)}
In other words, M 5-branes each contribute a fraction ${1\over 24}$ of the
total anomaly in Eq.\sixdanom\ above.

Similarly, this $U(1)$ is anomalous on an M-theory orientifold 5-plane, the
anomaly in this case being 
\eqn\fivepanom{
\Delta = {1\over 48}\int_{M^6} {F\over 4\pi}\wedge p_1(R)\, \Sigma(x)}
or a fraction ${1\over 96}$ of the total. The above two formulae added up
over 16 M 5-branes and 32 M orientifold 5-planes precisely reproduce
Eq.\sixdanom. Moreover, the M-theory relation with type IIA is satisfied,
with these anomalies being related to vacuum momentum and hence eventually
to gravitational WZ couplings. 

One might worry that in this case the relationship between $U(1)$ anomalies
and vacuum momentum is not so obvious, since the M 5-brane and orientifold
5-planes apparently have gravitational anomalies. However, as stressed in
Ref.\witmfive, gravitational anomaly inflow from the bulk actually renders
both of these objects anomaly-free separately.

The principal consequence of the above discussion is that the $U(1)$
anomaly of Ref.\refs\greengab\ is manifested in M-theory, but not in the
bulk (this would be impossible just because the bulk is non-chiral). It
appears on the chiral objects of the theory, namely 5-branes and
orientifold 5-planes.

\newsec{Conclusions}

We have shown that the $U(1)$ and $SL(2,Z)$ anomalies recently discussed in
Ref.\greengab\ are consistently connected to a number of stringy dualities
and even to M-theory. This is particularly satisfying since the original
derivation is based more directly on properties of  supergravity
than of string theory.

We have re-derived this anomaly using the fact that type IIB string theory
has a hidden ``vacuum momentum'' which in turn is predicted by T-duality
with type IIA theory. The relation of this vacuum momentum to the $U(1)$
anomaly is embodied in Eq.\reln\ above, along with the absence of
gravitational anomalies in 8-fold compactifications. 

It would be even more satisfying to have a direct derivation of Eq.\reln\
on general grounds. Each term is a kind of anomaly in a different symmetry:
$A_{U(1)}$ is of course the $U(1)$ anomaly, $A_{\rm grav}$ is the
gravitational anomaly and $A_{\rm vac}$, the vacuum momentum crucial to
this story, is roughly like an anomaly in the $U(1)$ rotation generated by
$L_0 - \barL_0$. 

The constraint that $\chi/24\in Z$ turns out to come from modular
invariance in this context. It can be modified by the inclusion of
background flux, which leads to the prediction that the $U(1)$ anomaly of
Ref.\greengab\ should be modified if the self-dual 4-form is turned on. 

Finally, it was observed that the M5-brane and M orientifold 5-plane
have vacuum momentum hidden in them by virtue of their chirality, this
fits in with nontrivial M-theory dualities and also with their direct
relation to 4-branes and 4-planes of type IIA string theory. In turn,
this phenomenon is related to $U(1)$ anomalies on the world-volumes of
these objects, similar to those found in 10 dimensional type IIB
theory in Ref.\greengab.

It may be hoped that these various facts suggest something about the
larger picture in which they fit together. For example, it is
intriguing that none of the M-theory defects (branes and planes) have
gravitational couplings of the type carried by D-branes (membranes,
like D2-branes, have too low a dimension to support gravitational
Chern-Simons-type couplings). The discussions above also illustrate
once more the general theme that branes and planes are similar in some
respects, the principal difference being that the latter have no
independent world-volume fields of their own.
\bigskip

\noindent{\bf Acknowledgements}

I am grateful to Rajesh Gopakumar, Albion Lawrence and Samir Mathur
for helpful discussions, and Ruben Minasian for a useful
correspondence. Hospitality and financial support from Andy
Strominger and the Physics Department at Harvard University are
gratefully acknowledged.

\listrefs \end